\documentclass[aps,preprintnumbers, superscriptaddress, amsmath, showpacs]{revtex4}
\usepackage{epsfig}
\usepackage{psfrag}

\usepackage{graphicx}
\usepackage{dcolumn}
\usepackage{bm}

\begin{document}

\preprint{RU06-9-B}

\title{Higher Angular Momentum Mixing in a Non-spherical Color
Superconductor with Time Reversal Invariance Violation}

\author{Bo  Feng}
 \email{fengbo@iopp. ccnu. edu. cn}
\affiliation{Institute of Particle Physics,  Huazhong Normal University,
Wuhan,  430079,  China}
\author{ De-fu  Hou}%
 \email{hdf@iopp. ccnu. edu. cn}
 \affiliation{Institute of Particle Physics,  Huazhong Normal University,
Wuhan,  430079,  China}
\author{Hai-cang Ren}
\email{ren@mail.rockefeller.edu} \affiliation{Physics Department, The
Rockefeller University, 1230 York Avenue,  New York,  NY
10021-6399} \affiliation{Institute of Particle Physics, Huazhong
Normal University, Wuhan,  430079,  China}

\date{\today}

\begin{abstract}
The angular momentum mixing in a non-spherical CSC with nonzero
azimuthal quantum number and therefore violating time reversal
invariance has been examined. The mixing is bound to occur because
of the equal strength of the pairing potential mediated by
one-gluon exchange for all partial waves to the leading order QCD
running coupling constant and the nonlinearity of the gap
equation. The free energy with mixing is lower than that with
$p$-wave pairing only, but still higher than that of the spherical
CSL state.
\end{abstract}

\maketitle

\section{Introduction}\label{sec:level1}
It has long been expected that quark matter at high baryon density
and low temperature becomes a color
superconductor(CSC)\cite{B,DA}. CSC is characterized by a diquark
condensate, which is analogous to the Cooper pair in an ordinary
superconductor but the structure of the condensate is much richer
because of the nonabelian color and flavor charges.

For very large baryon density, where the masses of u,d and s
quarks can be ignored, the ground state is in the
color-flavor-locked(CFL) phase\cite{MKF}. The situation becomes
more involved in moderate density because of the strange quark
mass, $\beta$ equilibrium and the charge neutrality conditions,
which will induce a substantial Fermi momentum mismatch among
different quark flavors and thereby reduce the available phase
space for Cooper pairing. A number of exotic CSC phases have been
proposed in the presence of mismatch, but a consensus point of
view of the true ground state has not been
reached\cite{MI,MCK,ABR,BCR}. An interesting alternative in this
circumstance is the single flavor pairing, which is obviously free
from the Fermi momentum mismatch. Since the attractive interaction
of quarks is provided in the antisymmetric color-antitriplet
channel, the total angular momentum channel must be symmetric in
order to insure the overall antisymmetric of the Cooper pair wave
function as required by Pauli principle. Therefore, Cooper pair in
single flavor pairing should be implemented at a higher total
angular momentum, the obvious choice is the $p$-wave pairing(J=1),
which had been extensively explored in the
literatures\cite{D,RD,T,A}.

The energy gaps in the single flavor pairing can be divided into
two categories, spherical gaps (e.g. CSL) and non-spherical ones
(e.g. polar phase and A phase). While a spherical gap is made of
$p$-wave alone, a non-spherical gap may contain higher partial
waves because of the nonlinearity of the gap equation. It was
argued in the literature that the contribution of higher partial
waves is of higher order in $g$ with $g$ the running coupling
constant of QCD. This, however, is not the case, as we shall
explain.

In a previous paper\cite{BDH}, we considered the single flavor CSC
with longitudinal pairing, in which the pairing two quarks have
equal helicity. Because of the equal strength of the pairing
potential mediated by one-gluon exchange for all partial waves to
the leading order QCD running coupling constant, a non-spherical
pairing receives contributions from all odd $J$'s at the
{\it{same}} order of $g$. A consistent solution to the gap
equation with a definite azimuthal quantum number $M$, however,
can be constructed. For the CSC that pairs red and green quarks
only, the energy gap with a definite $M$ is of the form
\begin{equation}
\Delta=e^{iM\varphi}\Delta_0f(\cos\theta) \label{cmplxgp}
\end{equation}
with $\theta$ and $\varphi$ the polar angles of the relative
momentum $\vec p$ of the pairing quarks, where
\begin{equation}
f(\cos\theta)=\sum_{J\ge |M|, J={\rm
odd}}b_JP_J^M(\cos\theta)\label{2}
\end{equation}
with $P_J^M(x)$ the associate Legendre function. The free energy
will be brought down by the mixing in comparison with that
reported in the literature. For $M=0$, we found that
\begin{equation}
f(\cos\theta)=e^{-17/3}[0.9866P_1(\cos\theta)-0.0673P_3(\cos\theta)+0.0214P_5(\cos\theta)+...].
\end{equation}
The magnitude of the condensation energy is raised by 3.5 percent,
which is too small to compete with the spherical
color-spin-locked(CSL) state in single CSC. As we pointed out in
\cite{BDHt}, the energy balance between the CSL and non-spherical
states maybe offset by the presence of anisotropy or the violation
of time reversal invariance. In a compact star, the magnetic field
and the stellar revolution provides such an opportunity.

In this paper, we extend the analysis in \cite{BDH} to case with a
nonzero azimuthal quantum number ($M\neq 0$). We will firstly
derive the nonlinear integral equation for the angular dependence
of the gap function and then give the numerical solution to this
equation. Finally, a summary and concluding remarks will be given.

\section{The integral equation of the angular dependence}
The non-zero azimuthal quantum number violate the time reversal
invariance and the gap function is proportional to $e^{i\varphi}$
in this case\cite{BDJH}. The condensation energy density for
single flavor CSC with longitudinal pairing reads\cite{BDH}
\begin{eqnarray}
\nonumber F=&-&\frac{3\bar{g}^2\mu^4}{32\pi^4}\int d\nu \int
d\nu^{\prime}\int d^2\hat{p}\int
d^2\hat{p^{\prime}}V(\nu-\nu^{\prime},\Theta)\times\frac{\phi(\nu,\hat{p})\phi(\nu^{\prime},\hat{p^{\prime}})}{\sqrt{[\nu^2+|\phi(\nu,\hat{p})|^2][\nu^{\prime2}+|\phi(\nu^{\prime},\hat{p^{\prime}})|^2]}}\\
&+&\frac{2\mu^2}{(2\pi)^3}\int d\nu\int
d^2\hat{p}\Big{[}|\nu|-\frac{\nu^2}{\sqrt{\nu^2+|\phi(\nu,\hat{p})|^2}}\Big{]}
\label{freeenergy}
\end{eqnarray}
where $\nu$ and $\nu^\prime$ are Euclidean energies, $\Theta$ is
the angle between $\hat p$ and $\hat p^\prime$ and V is the
pairing potential mediated by the one-gluon-exchange of QCD
\begin{equation}
V(\nu-\nu^{\prime},\Theta)=D_l(\nu-\nu^{\prime},\Theta)+D_t(\nu-\nu^{\prime},\Theta)
\end{equation}
with $D_l$ and $D_t$ is the longitudinal(color electric) and
transverse(color magnetic) part of hard-dense-loop(HDL) gluon
propagator respectively. The gap function $\phi(\nu,\hat p)$ is
extracted from the Nambu-Gorkov off diagonal blocks of the quark
self-energy and its value at $\nu=0$ gives $\Delta$ of
(\ref{cmplxgp}). The gap equation can be derived by minimizing the
condensation energy with respect to the gap function
\begin{equation}
\frac{\delta F}{\delta\phi}=0
\end{equation}
and we end up with
\begin{equation}
\phi(\nu,\hat{p})=\frac{g^2\mu^2}{24\pi^3}\int d\nu^{\prime}\int
d^2\hat{p}^{\prime}V(\nu-\nu^{\prime},\theta)\frac{\phi(\nu^{\prime},\hat{p}^{\prime})}{\sqrt{\nu^{\prime2}+|\phi(\nu^{\prime},\hat{p}^{\prime})|^2}}
\label{qcdgapeq}
\end{equation}
A consistent derivation of the gap equation up to the subleading
order need to include the one-loop self energy of quarks, the net
result is to replace the first term in the square root on RHS of
Eq.(\ref{qcdgapeq}) by $\nu^{\prime 2}/Z^2(\nu^{\prime})$ with
$Z(\nu^{\prime})$ the wave function renormalization
factor\cite{WJH,Qun}. But it will not interfere with the angular
dependence of the gap function as we have seen in\cite{BDH}.

The potential $V$ can be expanded in terms of Legendre
polynomials\cite{WJH}
\begin{equation}
V(\nu-\nu^{\prime},\Theta)=\frac{1}{6\mu^2}{\rm
ln}\frac{\omega_c}{|\nu-\nu^{\prime}|}\sum\limits_{J=0}^\infty(2J+1)P_J(\cos\Theta)
+\frac{1}{2\mu^2}\sum\limits_{J=1}^\infty (2J+1)c_JP_J(\cos\Theta)
\label{expansion}
\end{equation}
with $\omega_c=\frac{1024\sqrt{2}\pi^4\mu}{g^5}$ and $c_J$ given
by
\begin{equation}
c_J=-2\sum_{n=1}^{J}\frac{1}{n}.
\end{equation}
From this expansion, we can see that the pairing strength are
equal to leading order of the QCD running coupling constant, but
subleading terms fall off with increasing $J$. It is this falling
off that makes the amount of the angular momentum mixing
numerically small for the solutions in\cite{BDH} and in this
letter as we shall see below.

Substituting Eq. (\ref{expansion}) into the gap equation
(\ref{qcdgapeq}), we have
\begin{eqnarray}
\nonumber \phi(\nu,\hat
p)=&&\bar{g}^2\int\limits_{0}^{\omega_0}d\nu^{\prime}\Big\{\frac{1}{2}\big({\rm
ln}\frac{\omega_c}{|\nu-\nu^{\prime}|}+{\rm
ln}\frac{\omega_c}{\nu+\nu^{\prime}}\big)\frac{\phi(\nu^{\prime},\hat
p)}{\sqrt{\nu^{\prime 2}+|\phi(\nu^{\prime},\hat
p)|^2}}\\
&+&\frac{3}{2\pi}\int d\hat p^{\prime}\frac{1}{1-\hat p\cdot\hat
p^{\prime}}\Big[\frac{\phi(\nu^{\prime},\hat p^{\prime})}
{\sqrt{\nu^{\prime2}+|\phi(\nu^{\prime},\hat
p^{\prime})|^2}}-\frac{\phi(\nu^{\prime},\hat p)}
{\sqrt{\nu^{\prime2}+|\phi(\nu^{\prime},\hat p)|^2}}\Big]\Big\}
\label{gapequation}
\end{eqnarray}
where $\bar g^2=g^2/(18\pi^2)$ and a UV cutoff $\omega_0\sim g\mu$
is introduced. On writing
\begin{equation}
\phi(\nu,\hat p)=e^{iM\varphi}\psi(\nu,\cos\theta),
\end{equation}
the equation for the real function $\psi(\nu,\theta)$ reads

\begin{eqnarray}
\nonumber
\psi(\nu,x)=&&\bar{g}^2\int\limits_{0}^{\omega_0}d\nu^{\prime}\Big\{\frac{1}{2}\big({\rm
ln}\frac{\omega_c}{|\nu-\nu^{\prime}|}+{\rm
ln}\frac{\omega_c}{\nu+\nu^{\prime}}\big)\times\frac{\psi(\nu^{\prime},x)}{\sqrt{\nu^{\prime
2}+\psi^2(\nu^{\prime},x)}}\\&+&3\int_{-1}^{1} d
x^{\prime}\Big[\frac{\psi(\nu^{\prime},x^{\prime})I_M(x,x^\prime)}
{\sqrt{\nu^{\prime2}+\psi^2(\nu^{\prime},x^{\prime})}}
-\frac{1}{|x-x^{\prime}|}\frac{\psi(\nu^{\prime},x)}
{\sqrt{\nu^{\prime2}+\psi^2(\nu^{\prime},x)}}\Big]\Big\}
\label{gapequation}
\end{eqnarray}
where $x=\cos\theta$ and
\begin{equation}
I_M(x,x^\prime)=\frac{1}{2\pi}\int_0^{2\pi}
d\varphi^{\prime}\frac{e^{iM(\varphi^{\prime}-\varphi)}}{1-\hat
p\cdot\hat p^{\prime}}=\hbox{real}
\end{equation}
The most favored pairing channel corresponds to $|M|=1$ and there
is no difference between $M$ and $-M$. We only consider the case
of $M=1$ below. We have $I_1(x,x^\prime)={\mathcal
I}(x,x^\prime)/|x-x^{\prime}|$ with ${\mathcal I}$ given by
\begin{equation}
{\mathcal
I}(x,x^\prime)=\frac{1-xx^{\prime}-|x-x^{\prime}|}{\sqrt{(1-x^2)(1-x^{\prime
2})}}
\end{equation}

The gap equation (\ref{gapequation}) can be further simplified by
using the approximation in\cite{D}
\begin{equation}
{\rm ln}\frac{\omega_c}{|\nu-\nu^{\prime}|}\simeq{\rm
ln}\frac{\omega_c}{|\nu_>|}
\end{equation}
with $|\nu_>|={\rm max}(|\nu|,|\nu^\prime|)$. Introducing new
variables
\begin{equation}
\xi=\rm ln \frac{\omega_c}{\nu},\hspace{0.5cm} a=\rm
ln\frac{\omega_c}{\omega_0}\label{newvar}
\end{equation}
and writing $\psi(\nu,x)$ as $\psi(\xi,x)$, we have
\begin{equation}
\psi(\xi,x)=\xi\Phi(\xi,x)-\int_{a}^{\xi}d\xi^{\prime}\xi^{\prime}\frac{d\Phi}{d\xi^{\prime}}+3\int_{-1}^{1}dx^{\prime}\frac{\Phi(a,x^{\prime}){\mathcal
I}(x,x^\prime)-\Phi(a,x)}{|x-x^{\prime}|}\label{gapeqPhi}
\end{equation}
where $\Phi(\xi,x)$ is defined by
\begin{equation}
\Phi(\xi,x)\equiv\bar{g}^2\int_{\xi}^{\infty}d\xi^{\prime}\frac{\psi(\xi^{\prime},x)}
{\sqrt{1+\frac{\psi^2(\xi^{\prime},x)}{\omega_c^2}e^{2\xi^{\prime}}}}
\end{equation}
Taking the first derivative of both sides with respect to $\xi$ in
Eq.(\ref{gapeqPhi}), we have
\begin{equation}
\frac{d\psi(\xi,x)}{d\xi}=\Phi(\xi,x)\label{gapeq2}
\end{equation}
which implies
\begin{equation}
\frac{d\psi}{d\xi}\rightarrow 0 \label{boundary}
\end{equation}
as $\xi\rightarrow \infty$ for all $x$. Another derivative of
(\ref{gapeq2}) yields
\begin{equation}
\frac{d^2\psi(\xi,x)}{d\xi^2}+\frac{\bar{g}^2\psi(\xi,x)}{\sqrt{1+\frac{\psi^2(\xi,x)}{\omega_c^2}e^{2\xi}}}=0.
\label{gapeq}
\end{equation}
The $x$-dependence of $\psi(\xi,x)$ will be fixed by the
Eq.(\ref{gapeqPhi}) at $\xi=a$, i. e.
\begin{equation}
a\Phi(a,x)-\psi(a,x)+3\int_{-1}^{1}dx^{\prime}\frac{\Phi(a,x^{\prime}){\mathcal
I}-\Phi(a,x)}{|x-x^{\prime}|}=0 \label{43}
\end{equation}
The solution to (\ref{gapeq}) subject to the boundary condition
(\ref{boundary}) has been obtain in\cite{BDH} up to subleading
order, which applies to the case $M=1$ as well. We have
\begin{equation}
\psi(a,x)\simeq\Delta_0f(x)\Big{[}\frac{\pi}{2}-\bar g(b-a)-\bar
g{\rm ln}2\Big{]}+O(g)\label{phiax}
\end{equation}
and
\begin{equation}
\Phi(a,x)=\bar g\Delta_0f(x)+O(g)\label{Phia}
\end{equation}
where $\Delta_0$ is the s-wave gap given by
\begin{equation}
\frac{\pi}{2}-\bar g{\rm ln}\frac{2\omega_c}{\Delta_0}=0
\end{equation}
and $b=\ln\frac{\omega_c}{\Delta_0|f(x)|}$. Substituting
Eq.(\ref{phiax}) and Eq.(\ref{Phia}) into Eq.(\ref{43}), we derive
the integral equation for the angle dependent factor $f(x)$
\begin{equation}
f(x){\rm
ln}|f(x)|+3\int_{-1}^{1}dx^{\prime}\frac{f(x)-f(x^{\prime}){\mathcal
I}(x,x^\prime)}{|x-x^{\prime}|}=0 \label{56}
\end{equation}

\section{The solution to the integral equation}
As we have done in\cite{BDH}, this type of integral equation can
be solved by a variational method. Substituting the solution of
(\ref{gapeq}) and (\ref{boundary}) into (\ref{freeenergy}), the
condensation energy density becomes a functional of $f(x)$(see
appendix B in\cite{BDH}),
\begin{equation}
F=\frac{\mu^2\Delta_0^2}{2\pi^2}{\mathcal F}[f]
\end{equation}
with
\begin{eqnarray}
{\mathcal F}[f(x)]=\int_{-1}^{1}dxf^2(x)\Big{[}{\rm
ln}|f(x)|-\frac{1}{2}\Big{]}
+\frac{3}{2}\int_{-1}^{1}dx\int_{-1}^{1}dx^{\prime}\frac{f^2(x)-2f(x)f(x^{\prime}){\mathcal
I}+f^2(x^{\prime})}{|x-x^{\prime}|} \label{variational}
\end{eqnarray}
It can be easily verified that the variational minimization of
Eq.(\ref{variational}) does solve Eq.(\ref{56}). Before the
numerical calculations, we consider a trial function as
\begin{equation}
f(x)=cP_1^1(x)=c\sqrt{1-x^2}
\end{equation}

Substituting it into the condensation energy density
Eq.(\ref{variational}) and the minimization with respect to $c$
yields
\begin{equation}
c=\frac{1}{2}e^{5/6}e^{-6}
\end{equation}
at which
\begin{equation}
{\mathcal F}=-5.863\times 10^{-6}\label{initial}
\end{equation}
This trial function is what people carried over from $A$ phase in
${}^3\rm He$\cite{T,A}, which contains $p$-wave only, but it is
not optimal. The free energy will be lowered further by including
higher partial waves as we shall see.

\begin{figure}
\centering
\includegraphics[]{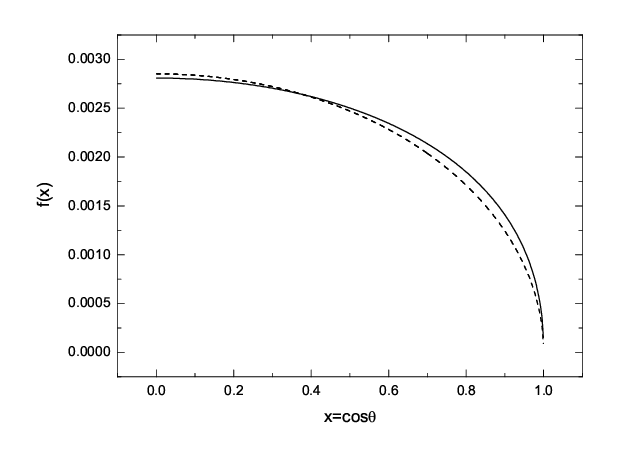}
\caption{\label{fig:eps1} The angular dependence of the gap
function with nonzero azimuthal quantum number. The dashed line
and solid one correspond to the trial function and the numerical
solution to Eq.(\ref{56}) respectively.}
\end{figure}

Following the same procedure in\cite{BDH}, we obtain the numerical
solutions to Eq.(\ref{56}). In Fig.\ref{fig:eps1}, we show the
angular dependence of the gap function with nonzero azimuthal
quantum number, the dashed line and the solid line are the trial
function and the numerical solution to Eq.(\ref{56}) respectively.
They depart from each other slightly, indicating a small mixture
of higher partial waves. We find the minimum value of the target
functional with the solid line
\begin{equation}
{\mathcal F}=-5.977\times 10^{-6}
\end{equation}
which drop from Eq.(\ref{initial}) by $1.9$ percent.

As we have found in\cite{BDH}, the drop of the condensation energy
with longitudinal pairing with zero azimuthal quantum number by
angular momentum mixing is numerically small. Here, with nonzero
azimuthal quantum number, the situation is also true. With these
small drop amount in condensation energy, the non-spherical
pairing in single flavor CSC can not compete with the spherical
pairing state CSL.

Regarding the angular momentum contents for our solutions
according to Eq.(\ref{2}) for $M=1$, we found
\begin{equation}
f(\cos\theta)=\frac{1}{2}e^{5/6}e^{-6}[1.0112P_1^1(\cos\theta)+0.0212P_3^1(\cos\theta)
+0.004P_5^1(\cos\theta)+...].
\end{equation}

\section{Concluding remarks}
In summary, we have explored the angular momentum mixing in a
non-spherical CSC with nonzero azimuthal quantum number, which
violates the time reversal invariance because the gap function is
not real in this case. The mixing is driven by the equal strength
of the pairing potential mediated by one-gluon exchange for all
partial waves to the leading order QCD running coupling constant
and the nonlinearity of the gap equation. However, the gaining
factor of the condensation energy caused by mixing is smaller than
that by forming the spherical CSL state. The angular momentum
mixing in the parallel case with transverse pairing has also been
considered in \cite{BDHt}, in which the amount of the free energy
drop is also small. Therefore, we conjecture that angular momentum
mixing of various non-spherical CSC is not sufficient to compete
with the CSL energetically in the ultra relativistic limit. A
rigorous proof of this statement is anticipating.

\begin{acknowledgments}

We would like to extend our gratitude to D. Rischke, T.
Sch$\ddot{a}$fer, A. Schmitt
 for stimulating discussions. We are also
benefitted from conversations with  J.R. Li and Q. Wang. The work
of D. F. H. and H. C. R. is supported in part by NSFC under grant
No. 10575043 and by US Department of Energy under grants
DE-FG02-91ER40651-TASKB. The work of D. F. H. is also supported in
part by Educational Committee under .This work was supported by
MOE of China under grant NCET-05-0675 project No. IRT0624

\end{acknowledgments}

\end{document}